\documentclass[a4paper,11pt]{article}
\pdfoutput=1 

\usepackage{jcappub} 

\usepackage[T1]{fontenc} 

\usepackage[utf8]{inputenc}

\usepackage[binary-units]{siunitx}
\DeclareSIUnit\parsec{pc}
\DeclareSIUnit\gauss{G}

\newcommand{\kpc}{\kilo\parsec}
\newcommand{\Mpc}{\mega\parsec}

\usepackage{pseudo}

\usepackage{cleveref}
\crefname{listing}{Algorithm}{Algorithms}  
\Crefname{listing}{Algorithm}{Algorithms}
\crefname{algorithm}{algorithm}{algorithms}  
\Crefname{algorithm}{Algorithm}{Algorithms}

\title{Static Multiresolution Grids with Inline Hierarchy Information for Cosmic Ray Propagation}
\author{Gero Müller}
\affiliation{III. Physikalisches Institut A, RWTH Aachen University, D-52056 Aachen, Germany}
\emailAdd{gero.mueller@physik.rwth-aachen.de}
\abstract{For numerical simulations of cosmic-ray propagation fast access to static magnetic field data is required.
We present a data structure for multiresolution vector grids which is optimized for fast access, low overhead and shared memory use. The hierarchy information is encoded into the grid itself, reducing the memory overhead.
Benchmarks show that in certain scenarios the differences in deflections introduced by sampling the magnetic field model can be significantly reduced when using the multiresolution approach.
}


\begin{document}
\maketitle
\flushbottom

\section{Introduction}

In astroparticle simulations charged particles, electrons, protons or ionized nuclei, are propagated through static galactic and extragalactic magnetic fields.
The magnetic field information can be extracted from large scale structure simulations, e.g. \cite{Sigl2003, Dolag2005}.
Modern magnetohydrodynamic (mhd) simulations are represented by so-called smooth particles, providing highly variable spatial resolutions.
Spatial properties are not stored in simple grids but by movable particles of spherical shape.
The radii of these spheres are defined by the distances to their neighboring particles.
Inside each sphere the cosmological properties like mass density, temperature and pressure are distributed following a kernel function.
The kernel used in \cite{Dolag2008} is:
\begin{equation}
   W(x,h)=\frac{8}{\pi h^3}\left\{\begin{array}{ll}
      1 - 6 \left(\frac{x}{h}\right)^2 + 6 \left(\frac{x}{h}\right)^3 \;\;& 0 \le \frac{x}{h} \le 0.5 \\
      2 \left(1 - \frac{x}{h}\right)^3                              & 0.5 \le \frac{x}{h} \le 1 \\
      0                                                             & 1 \le \frac{x}{h} \\
   \end{array} \right. , \label{SPH:kern}
\end{equation}
where $x$ is the distance to the center of the particle and $h$ is the radius of the sphere, the smoothing length. 
Accessing individual magnetic field vectors through on-the-fly calculation is computationally expensive, since they need to be calculated from many smooth particles using the kernel function.
To keep the high resolution when precalculating the magnetic field vectors, a multiresolution approach is required.
Volumes with almost equal magnetic field, mainly found in voids or filaments, are covered by only a few smooth particles with large radii.
These would be represented by large, low resolution grid cells.
On the other hand, highly detailed volumes in galaxy clusters are represented by small, high resolution grid cells.

To store the magnetic field of the local universe with an extent of \SI{240}{\Mpc} and a resolution of about \SI{50}{\kpc} using a regular grid would require roughly $4096^3$ samples, or \SI{768}{\gibi\byte}.
Using octrees \cite{Meagher1980}, k-d trees \cite{Bentley1975} or other spatial partitioning algorithms which are stored separately from the data would introduce significant overhead for very large fields.

The propagation codes CRT \cite{Sutherland2010} and CRPropa3 \cite{Batista2016} both feature composed magnetic fields, where a magnetic field  from a large scale structure simulation or analytical model \cite{Jansson2012, Pshirkov2011} is combined with a periodically repeated small scale random component.
The data structure presented in this paper can either be used to sample the analytic magnetic field or to use magnetic fields from other sources like magnetohydrodynamic simulations.
In both cases it improves the performance while retaining a high degree of detail. It can still be combined with a periodically repeated small scale random component.

We find the following requirements for the magnetic field data structure and algorithm:
a) Fast access to individual samples.
Cosmic rays travel a large distance and pass a large amount of magnetic field vector.
The simulation time significantly depends on the time it takes to retrieve a single magnetic field sample.
b) Often only a small subset of the magnetic field is required for a simulation.
In a typical simulation setup, cosmic rays are emitted by one source and detected at our Galaxy.
In this scenario only the field between the source and the observer is of interest.
The data structure should provide the means to dynamically load only the required samples from the disk into memory.
c) Cosmic ray propagation can be trivially parallelized since there are no interactions between the cosmic rays during propagation.
The data structure should be accessible from multiple simulation processes at the same time using memory mapped files.

In the next section we present a data structure for a multiresolution grid that fulfills all those requirements.
A brief overview on how to create such a multiresolution field is given in \cref{sec:creation}.
In \cref{sec:benchmark} we compare the performance to a simple cartesian grid when using up to 32 threads. The effect of different magnetic field resolutions on cosmic ray deflections is analyzed in \cref{sec:effect}.

\section{Data Structure} \label{sec:structure}

\begin{figure*}[htbp]
  \centering
  \includegraphics[width=\textwidth]{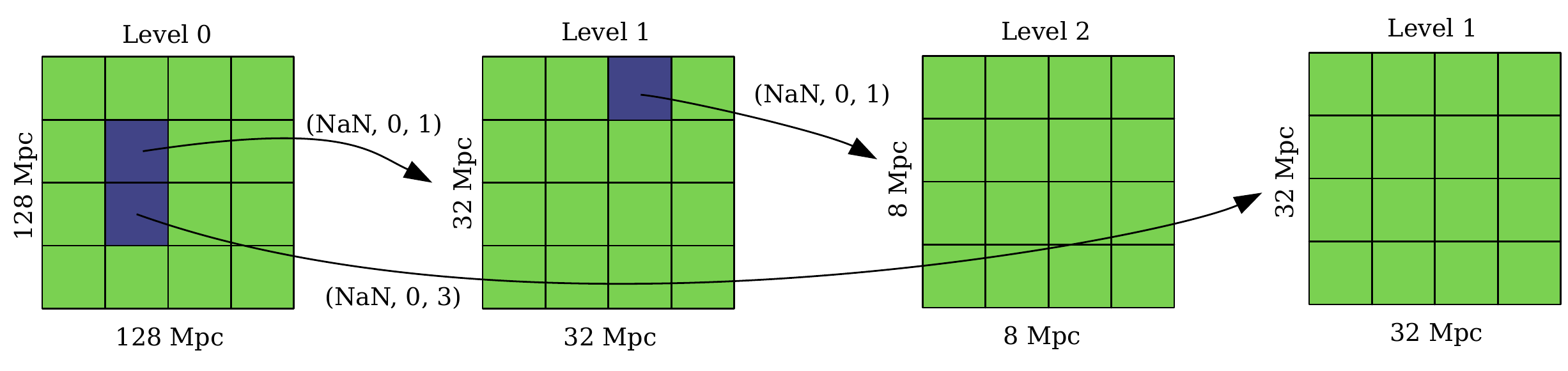}
  \caption{2D schematics of the hierarchical data structure. Bright cells contain actual field samples while dark cells contain relative indices to higher resolution grids.}
  \label{fig:hcube}
\end{figure*}

The multiresolution vector grid is stored in hierarchic cubes (HCube), outlined in \cref{fig:hcube}.
The top-most HCube in the hierarchy has the lowest resolution.
It has the index 0 and is located at the beginning of the memory block.
Each cell of the HCube contains either the sample for this position, or a relative index to the higher resolution HCube.
This index is relative to the referencing HCube.
References are encoded using NaN (not a number) for the x component of the vector while the y and z components contain the 64bit offset to the next HCube.
The memory address is then
\begin{equation}
  addr(next) = addr(this) + sizeof(HCube) * \textsl{uint64}(y, z)
\end{equation}
\textsl{uint64} converts two \SI{32}{\bit} floats to one \SI{64}{\bit} unsigned integer.
This way the tree structure is embedded into the data itself, contrary to other data structures like octrees or k-d trees.
Other data types than three dimensional vectors are possible as well, provided that the relative index can be encoded.
HCubes do not know their origin or size to reduce the memory overhead.
This requires that all HCubes have the same size.
The algorithm to access the samples of an HCube structure is given in \cref{alg:hcube_value}.

\begin{algorithm}[caption={Pseudocode for the lookup of a sample from an HCube. $cube\_cells$ is an array of 3D vectors with x, y, z components, $cube\_size$ is the side length of the HCube in physical units, $relative\_position$ is the position of the requested sample relative to the origin of the HCube in physical units, $cell\_count$ is the number of cells each HCube has per axis.}, label={alg:hcube_value}]
function sample (cube_cells, cube_size, relative_position, cell_count)
  cell_size := cube_size / cell_count
  indices := floor(relative_position / cell_size)

  if indices.x >= cell_count or
     indices.y >= cell_count or
     indices.z >= cell_count then
    error: Invalid position
  end

  index := 1D index from 3D indices
  cell := cube_cells[index]
  
  if cell.x = NaN then
    relative_index := integer from floats: cell.y, cell.z
    next_cube_cells := addr(this) + sizeof(this) * relative_index

    origin_of_this_cube := indices * cell_size
    next_relative_position := relative_position - origin_of_this_cube

    return sample(next_cube_cells, cell_size, next_relative_position, cell_count)
  else
    return cell
  end
end
\end{algorithm}

If the cells are stored in native binary format, they can be accessed using a memory mapped file allowing only partial reads and shared memory access from multiple processes.
The complexity of each access is of the order $\mathcal{O}(log_c(n))$ with c being the edge size of the HCubes and n being the edge size of the original field.
\Cref{table:overhead} shows the maximum overhead, i.e. the amount of extra memory needed if the whole volume is sampled in the highest possible resolution, for different HCube cell counts compared to a regular grid with the same resolution.
The overhead does not depend on the resolution, but only on the number of cells per axis per HCube.

\begin{table}[h]
  \centering
  \begin{tabular}{|cccc|}
    \hline
    \textbf{2} & \textbf{4} & \textbf{8} & \textbf{16} \\
    \hline
    14.29\%    & 1.59\%     & 0.20\%     & 0.02\% \\ 
    \hline
  \end{tabular}
  \caption{The maximum amount of overhead for HCube cell counts from 2 to 16 for any field.}
  \label{table:overhead}
\end{table}

\section{Creation} \label{sec:creation}

The HCube hierarchy can be created from any vector field, although for a sufficiently complex field the
gain compared to a cartesian grid may be minimal.
First, virtual memory for the maximum number of HCubes is reserved.
Then, each HCube in the tree is visited in a depth-first order.
New HCubes are allocated at the end of the current list of HCubes.
When the HCube with the maximum resolution is reached, the values of all cells are calculated.
Afterwards all values are evaluated and it is decided if the HCube can be collapsed and be described by its mean vector $\langle\vec{B}\rangle$.
The following rules apply: a) There must be no references among the cells b) All values must be sufficiently similar, and one of the following conditions must be fulfilled\footnote{These conditions are currently implemented, but might be replaced by more advanced conditions in the future.}:
\begin{eqnarray}
  \max(|\vec{B}_i - {\langle\vec{B}\rangle}|) &<& \epsilon \cdot |\langle\vec{B}\rangle| \\
  \max(|\vec{B}_i - {\langle\vec{B}\rangle}|) &<& \delta
\end{eqnarray}
$\vec{B}_i$ are the cell values, $\epsilon$ the allowed relative error and $\delta$ the allowed absolute error.
Finally, when all allocated HCubes have been visited, the file associated with the virtual memory is truncated after the last HCube.

A cartesian grid of single precision three vectors with 4096 samples per side requires \SI{768}{\gibi\byte}, for example.
It can be represented with 6 layers of HCubes with 4 samples per side.
The resolution is then given by $extent / 4096$.
A larger number of samples per HCube improves the performance since fewer layers need to be traversed.
Fewer layers on the other hand increase the space requirements, since fewer HCubes can be collapsed, i.e. represented by a single value.

\section{Benchmark} \label{sec:benchmark}

To benchmark the data structure and algorithm described above we use the open source C++ reference implementation \textit{quimby}\footnote{\url{https://forge.physik.rwth-aachen.de/public/quimby}}.
Quimby is a software library to read and process GADGET \cite{Dolag2008} files containing smooth particles from mhd simulations and it contains different algorithms and data structures to handle large magnetic field data.
Based on this library various tools for the creation of magnetic fields in different formats and from different sources are implemented.
Additionally Python bindings are provided, so it can easily be used together with CRPropa3.

\begin{figure}
  \flushleft
  \begin{minipage}[t]{3in}
    \includegraphics[width=3in]{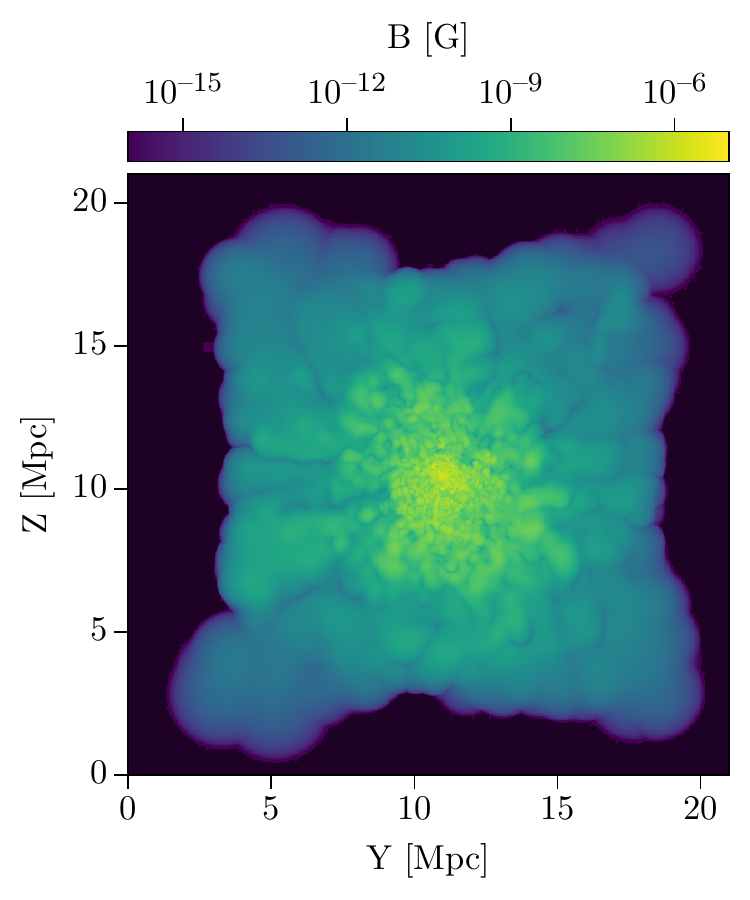}
  \end{minipage}\hfill
  \begin{minipage}[t]{3in}
    \includegraphics[width=3in]{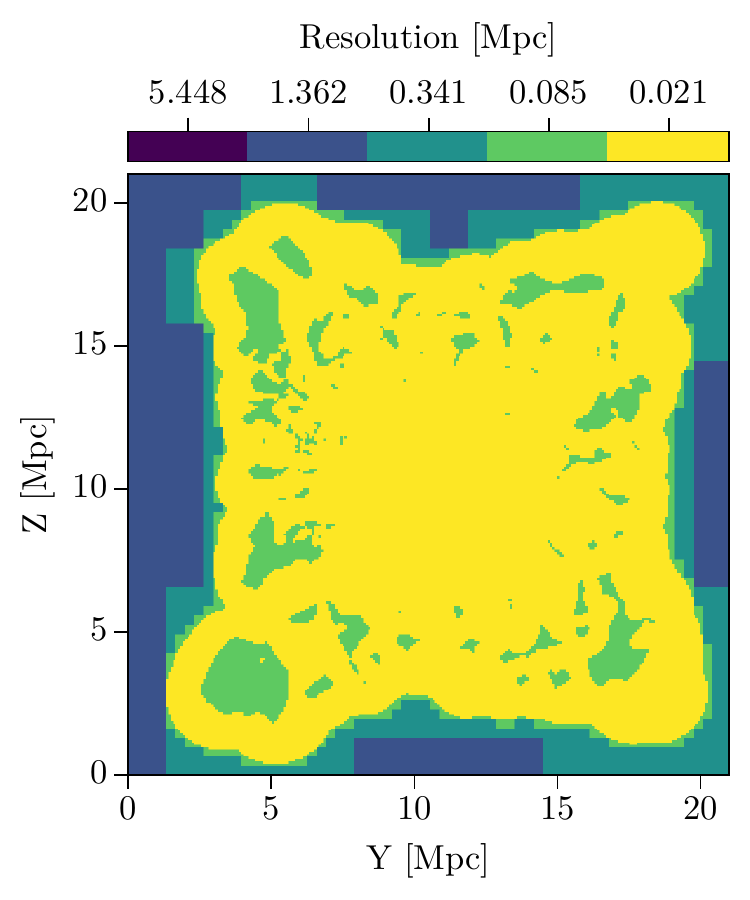}
  \end{minipage}
  \caption{Cross section of the magnetic field strength (left) and resolution (right) in the vicinity of a galaxy cluster. Created from HCubes with a size of 4, a relative error of $\epsilon$ = 0.1 and an absolute error of $\delta$ = \SI{e-19}{\gauss}.}
  \label{fig:coma}
\end{figure}

For this benchmark we use snapshots of mhd simulations of large-scale structure formations \cite{Dolag2005} to benchmark the look-up performance of our algorithm.
It consists of 317349 smooth particles in a box of 21$^3$ Mpc$^3$ and represents a subset of the magnetic field in the vicinity of a galaxy cluster as shown in \cref{fig:coma}.

First, a create cartesian grid with 1024 samples per axis is created from the smooth particles .
Based on this grid a multiresolution field is created with a HCube size of 4, a relative error of $\epsilon$ = 0.1 and an absolute error of $\delta$ = \SI{e-19}{\gauss}.
The final resolution for each pixel is shown for a slice in \cref{fig:coma}.
While the cartesian grid with 1024 samples requires 12 GiB the HCube grid only requires 5.5 GiB.

The performance of the regular grid and the HCube implementation is measured by two different access patterns.
The first pattern is a random access pattern where a random position inside the simulation volume is chosen.
This stresses the CPU caches, as no consecutive request can be served from cached memory pages.
The second pattern is a random-walk pattern, where the next position is chosen randomly within \SI{36}{\kpc} from the current position, allowing some requests to be served from memory pages still in the CPU caches. 

In a typical simulation both access patterns occur.
Cosmic rays with high rigidities have large step sizes, producing a random access like pattern.
In strong magnetic fields or when the comic ray rigidity is low, the magnetic field is queried in a pattern similar to a random walk.
For a \SI{100}{\Mpc} long trajectory of a a 50 EeV protons roughly $10^3 - 10^4$ magnetic fields vectors are queried.

All benchmarks are also performed using a null field implementation which simply returns a fixed sample, to measure the overhead of the benchmark.
We perform the benchmarks with the two access patterns on the regular grid, the HCube and the null field using between 1 and 32 threads.
Every run simulates $10^8$ field accesses.
We repeat each run 10 times and pick the fastest one, to discard slowdowns triggered by other activities on the computer. 

\begin{figure}
  \centering
  \begin{minipage}[c]{3in}
    \includegraphics[width=3in]{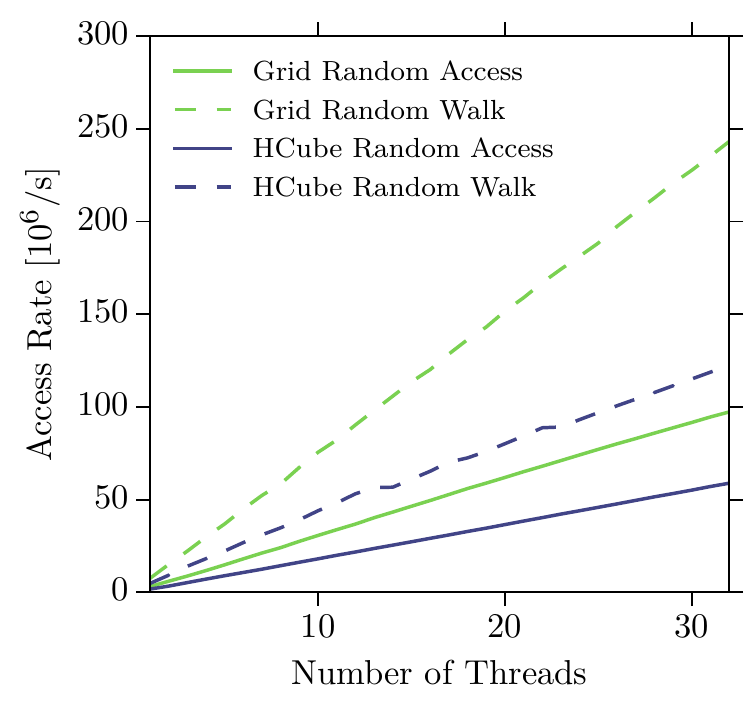}
    \caption{Benchmark results for a grid (HCube) with a (maximal) resolution of \SI{20.5}{\kpc} (\SI{21}{\Mpc}/1024)}
    \label{fig:1024-benchmark}
  \end{minipage}
\end{figure}

In \cref{fig:1024-benchmark} we show the results of the benchmark executed on a Intel(R) Xeon(R) X7550 @ 2.00GHz CPU.
The overhead, estimated by the null field, is subtracted. 
For the random-walk access pattern the multiresolution grid access rate is roughly faster by a factor of 2.
When accessing the field randomly it is only 1.6 times faster. 
The access rates for all patterns and implementations scale linearly with up to 32 threads.
Small deviations from linearity might result from environmental effects, e.g. other activities on the computer, or technical reasons like limited cache or memory bandwidth or even the CPU affinity of threads.

\section{Effect on Deflections} \label{sec:effect}

In this section we analyze the effect of different magnetic field sampling resolutions on the deflection of cosmic rays.
We again use the magnetic field from a large scale structure simulation \cite{Dolag2005}. It has an extent of \SI{240}{\Mpc} and a resolution up to \SI{10}{\kpc}.

The field is sampled with a resolution of \SI{14.7}{\kpc} (\SI{240}{\Mpc} with 16384 samples) and converted to three HCube fields with relative errors $\epsilon$ of 0.9, 0.4 and 0.1 and an absolute error $\delta$ of $10^{-14}$.
For comparison a cartesian grid with a resolution of \SI{234.4}{\kpc} (1024 samples) is included in the studies.
Each sample in this lower resolution field is the mean of 4096 values with \SI{14.7}{\kpc} resolution.
For more efficient ways to calculate low resolution fields from smooth particles see e.g. \cite{Pakmor2006}.
The low resolution field and the HCube field with $\epsilon = 0.4$ has a size of roughly 12 GiB, while the field with $\epsilon = 0.9$ is ten times smaller and the more detailed field with $\epsilon = 0.1$ is ten times larger.

The study is conducted in three regions:
in the void and filaments, at the border of a galaxy cluster and inside a galaxy cluster.
In the void and filament region large areas show a  very coherent field, which is an artifact of this large scale structure simulation and its seed field.
The border region contains areas with high and low resolution, while the region around the center of the cluster contains volumes of very high resolution and larger magnetic fields.
Each region is confined to a sphere of \SI{5}{\Mpc} radius.

\begin{figure*}[htbp]
  \includegraphics[width=6.5in]{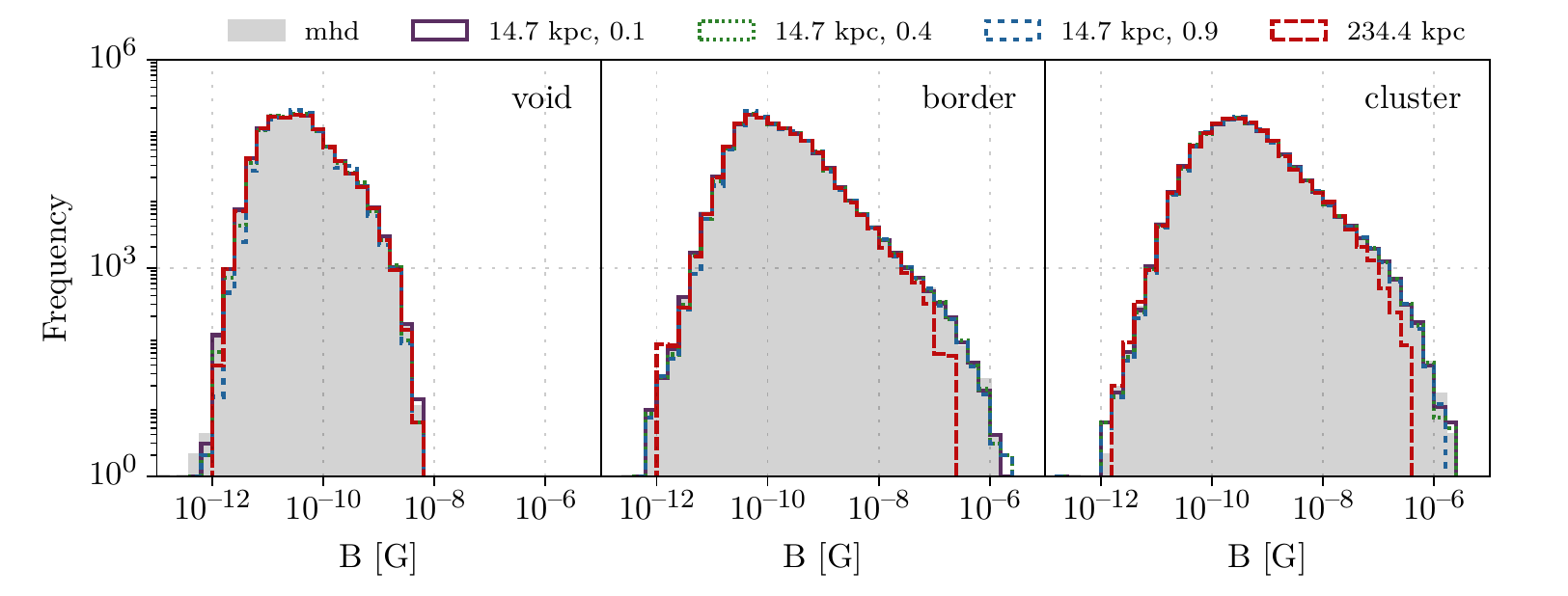}
  \caption{Distribution of $10^6$ magnetic field values in the void, at the border and inside a galaxy cluster.}
  \label{fig:field}
\end{figure*}

In \cref{fig:field} the distributions of $10^6$ magnetic field values from random positions inside each sphere are shown.
Large deviations from the original distribution (mhd) become evident only at the lowest resolution.
The median magnetic field strengths are approximately $10^{-10.5}$ G in the void, $10^{-10.0}$ G in the border and $10^{-9.5}$ G in the central region.
To estimate the effect on the deflection of cosmic rays we propagate isotropically emitted protons from the center to the border of the sphere using CRPropa3.
The same $10^6$ initial cosmic rays are propagated through the unsampled (mhd) and all four sampled fields.
The angular distance of the piercing point to the piercing point of the unsampled (mhd) magnetic field is finally recorded.
In the simulation interactions with the extragalactic background light (EBL) are disabled so only deflections in the magnetic fields are attended.
The lower limit of the integration step is \SI{0.5}{\kpc}, the upper limit is \SI{50}{\kpc} and the limit  for relative  error is $10^{-4}$.
Proton energies have been chosen to produce a median deflection of \SI{20}{\degree} in each scenario with the unsampled magnetic field:
0.16 EeV in the void, 1.8 EeV at the border and 45 EeV in the cluster.

\begin{figure*}[htbp]
  \includegraphics[width=6.5in]{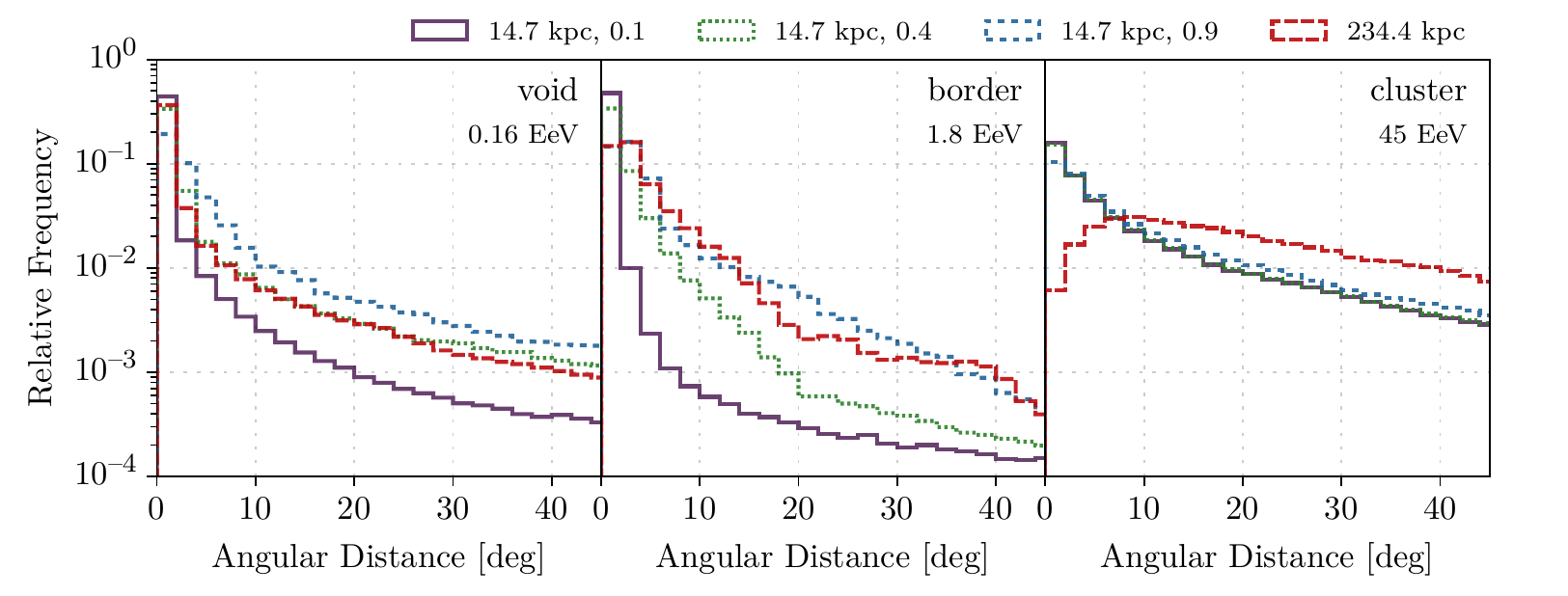}
  \caption{Distribution of the angular distances between the piercing points for different magnetic field resolutions in the void, at the border and inside a galaxy cluster.}
  \label{fig:angle}
\end{figure*}

The resulting distributions are shown in \cref{fig:angle} and the median values are tabulated in \cref{table:median}.
In the void the higher resolution does not improve the simulation results; only the very detailed model ($\epsilon = 0.1$) shows a slightly lower tail.
In the border region the level of detail in the high resolution fields has a stronger effects for some trajectories, while the high resolution field with the least level of detail ($\epsilon = 0.9$) performs comparably to the lower resolution field. 
A different picture presents itself in the cluster region.
All higher resolution fields perform better than the low resolution field, regardless of the level of detail.

Although the overall deviations from the unsampled magnetic field (mhd) can be large in certain scenarios, the use of the multiresolution HCube algorithm improves the simulation results in regions with large field strengths and small scale structures. The median angular distance can be reduced from above \SI{25}{\degree} to below \SI{5}{\degree} inside galaxy clusters.

\begin{table}[h]
  \centering
  \begin{tabular}{|cr|cccc|}
    \hline
    && \textbf{14.7 kpc} & \textbf{14.7 kpc} & \textbf{14.7 kpc} & \textbf{234.4 kpc} \\
    && $\epsilon=$ \textbf{0.1} & \textbf{0.4} & \textbf{0.9} & - \\
    \hline
    \textbf{void} & 0.16 EeV & 0.1$^\circ$ &  1.0$^\circ$ &  2.9$^\circ$ &  0.4$^\circ$ \\
    \textbf{border} & 1.8 EeV & 0.2$^\circ$ &  1.2$^\circ$ &  3.2$^\circ$ &  1.9$^\circ$ \\
    \textbf{cluster} & 45 EeV & 4.6$^\circ$ &  4.8$^\circ$ &  6.9$^\circ$ &  25.3$^\circ$ \\
    \hline
  \end{tabular}
  \caption{Median angular distance between the piercing points of the unsampled and sampled magnetic fields.}
  \label{table:median}
\end{table}

Finally we compare the runtime of the different implementations, presented in \cref{table:runtime}.
Different to the benchmarks in \cref{sec:benchmark} these runtime values are influenced by the following effects:
a) Especially in the cluster region the HCube hierarchy is deeper and more cell lookups are required.
b) Larger field strengths and small scale structures in the cluster region also require smaller step sizes in the integration code to satisfy the mandated accuracy.
c) Larger deflections induce an elongation of the trajectory length.

The unsampled field where the smooth particles are queried directly is much slower than the sampled fields.
Although the implementation for the direct access is not optimal, potential improvements will unlikely reduce the numbers significantly.
In the cluster region, where the improvements in quality are best, the runtime increases by a factor of 5 to 6.
In the border and void region the runtime increases only by a factor of 2.

\begin{table}[h]
  \centering
  \begin{tabular}{|cr|ccccc|}
    \hline
                     &          & \textbf{14.7 kpc} & \textbf{14.7 kpc} & \textbf{14.7 kpc} & \textbf{234.4 kpc} & \textbf{mhd} \\
                     &          & $\epsilon=$ \textbf{0.1} & \textbf{0.4} & \textbf{0.9} & - & - \\
    \hline
    \textbf{void}    & 0.16 EeV & 2  &  2 &  2 &  1 & 11 \\
    \textbf{border}  & 1.8 EeV  & 3  &  2 &  2 &  1 & 16 \\
    \textbf{cluster} & 45 EeV   & 6  &  5 &  5 &  1 & 36 \\
    \hline
  \end{tabular}
  \caption{Runtime of the benchmarks, relative to the lowest resolution (\SI{234.4}{\kpc}) in each scenario.}
  \label{table:runtime}
\end{table}

\section{Conclusion}

Magnetic fields play a vital role in cosmic ray propagation.
The deflection of charged particles changes not only the arrival directions but can also cause significant increases in the trajectory length, and therefore change the observed chemical composition.
The stochastic nature of the interactions with the EBL require time consuming simulations, so efficient access to magnetic fields are of great importance, especially with upcoming magnetic fields from larger and higher resolution large scale structure simulations (Dolag et. al, in prep.)\footnote{\href{http://www.magneticum.org}{http://www.magneticum.org}}.

We present a multiresolution grid optimized for fast access to large static vector fields.
Using this data structure it is possible to better utilize high resolution magnetic fields in cosmic ray propagation simulations.
Our benchmarks show that it performs well, is suitable to be used in parallel computing, and reduces the error of deflections induced by the sampling of the magnetic fields.

\acknowledgments

We are grateful for financial support to the Ministerium für Wissenschaft und Forschung des Landes Nordrhein-Westfalen, and to the Bundesministerium für Bildung und Forschung (BMBF).

\bibliographystyle{ieeetr}

\end{document}